\documentclass[showpacs,amsmath,amssymb,aps,showkeys,floatfix,prd,a4paper]{revtex4}

\usepackage{dcolumn}
\usepackage{bm}
\usepackage{epsfig}
\usepackage{amsfonts}
\usepackage{amssymb,amscd}

\def\lsim{\raise0.3ex\hbox{$<$\kern-0.75em\raise-1.1ex\hbox{$\sim$}}}

\def\gsim{\raise0.3ex\hbox{$>$\kern-0.75em\raise-1.1ex\hbox{$\sim$}}}

\newcommand{\be}{\begin{equation}}

\newcommand{\ee}{\end{equation}}

\def\beq{\begin{equation}}

\def\eeq{\end{equation}}

\def\beqa{\begin{eqnarray}}

\def\eeqa{\end{eqnarray}}

\newcommand{\ba}{\begin{eqnarray}}

\newcommand{\ea}{\end{eqnarray}}

\def\gappeq{\mathrel{\rlap {\raise.5ex\hbox{$>$}}

{\lower.5ex\hbox{$\sim$}}}}

\def\lappeq{\mathrel{\rlap{\raise.5ex\hbox{$<$}}

{\lower.5ex\hbox{$\sim$}}}}

\def\Toprel#1\over#2{\mathrel{\mathop{#2}\limits^{#1}}}

\begin{document}

\title{Tau polarization in neutrino - nucleus interactions at the LHC energy range}

\author{Reinaldo {\sc Francener}}
\email{reinaldofrancener@gmail.com}
\affiliation{Instituto de Física Gleb Wataghin - UNICAMP, 13083-859, Campinas, SP, Brazil. }

\author{Victor P. {\sc Gon\c{c}alves}}
\email{barros@ufpel.edu.br}
\affiliation{Institute of Physics and Mathematics, Federal University of Pelotas, \\
  Postal Code 354,  96010-900, Pelotas, RS, Brazil}
\affiliation{Institute of Modern Physics, Chinese Academy of Sciences,
  Lanzhou 730000, China}

\author{Diego R. {\sc Gratieri}}
\email{drgratieri@id.uff.br}
\affiliation{Escola de Engenharia Industrial Metal\'urgica de Volta Redonda,
Universidade Federal Fluminense (UFF),\\
 CEP 27255-125, Volta Redonda, RJ, Brazil}
\affiliation{Instituto de Física Gleb Wataghin - UNICAMP, 13083-859, Campinas, SP, Brazil. }

\begin{abstract}
Considering that the study of neutrino - nucleus interactions with incident neutrino energy ranges in the GeV -- TeV range is feasible at the Large Hadron Collider, we investigate in this paper  the degree of polarization ${\cal{P}}$ of the (anti) tau lepton produced in (anti) tau neutrino - tungsten interactions. We estimate the  differential cross-sections and the longitudinal and transverse components of the tau lepton polarization as a function of the tau lepton energy and distinct values of the scattering angle, assuming different values for the energy of the incoming (anti) tau neutrino. Different models for the treatment of the nuclear effects in the parton distribution functions are assumed as input in the calculations. Our results indicate that $ {\cal{P}} < 1$ for the neutrino energies reached at the LHC and are almost insensitive to the nuclear effects. 
 
\end{abstract}

\pacs{12.38.-t, 24.85.+p, 25.30.-c}

\keywords{}

\maketitle

\vspace{1cm}

\section{Introduction}

Over the last years, tau neutrinos have been detected by Super-Kamiokande and IceCube observatories, considering its measurements of atmospheric and astrophysical neutrino data (For a recent review see Ref. \cite{MammenAbraham:2022xoc}). In particular, the astrophysical neutrinos observed in the TeV to PeV energy range are important tests of Standard Model Physics and probes of New Physics scenarios in an energy range well beyond the center-of-mass energies of current terrestrial experiments. More recently, TeV-energy neutrinos, have been measured at the Large Hadron Collider (LHC) by the FASER \cite{FASER:2021mtu,FASER:2023zcr} and SND@LHC \cite{SNDLHC:2023pun} experiments, and two orders of magnitude higher statistics are expected during the high-luminosity LHC (HL-LHC) era using the detectors that have been proposed to be housed in the Forward Physics Facility (FPF) \cite{Anchordoqui:2021ghd,Feng:2022inv}.  Such future data will allow us to investigate several aspects of tau neutrino physics in an unexplored energy range (For related studies, see e.g. Refs. \cite{Candido:2023utz,Cruz-Martinez:2023sdv}).

In this paper, we investigate the degree of polarization of the tau lepton produced in tau neutrino - nucleus ($\nu_{\tau} A$) interactions at the TeV energy range. Our study is strongly motivated by the analysis performed in Ref. \cite{Hagiwara:2003di}, which has demonstrated that the produced $\tau$'s have high degree of polarization and their spin direction depends non-trivially on the energy and the scattering angle of $\tau$ in the laboratory frame. As the tau immediately decays after its production, it is detected through its decay particle distributions, which are strongly dependent on the $\tau$ spin polarization. Therefore, a precise determination of the degree of polarization is fundamental for the reconstruction of the tau events. In recent years, some studies have analyzed the tau polarization in $\nu A$ interactions considering distinct theoretical approaches to treat the contributions associated with the quasi - elastic, resonance production and deep inelastic scattering (DIS) processes \cite{Hernandez:2022nmp,Zaidi:2023hdd,Isaacson:2023gwp}. Such analyzes have focused on the GeV energy range, which will be probed by the DUNE neutrino oscillation experiment \cite{DUNE:2020lwj,DUNE:2020ypp}. Our goal is to complement these previous studies, by providing the predictions for the differential cross-sections and the longitudinal and transverse components of the tau lepton polarization derived considering   $\nu_{\tau} A$ interactions in the TeV energy range that will be probed at the LHC (For similar studies in the PeV energy see e.g. Refs. \cite{Payet:2008yu,Arguelles:2022bma}). In our analysis, we will estimate these quantities assuming that the main contribution comes from the DIS process and taking into account of the nuclear effects in the parton distribution functions (PDFs). In particular, we will compare the predictions derived using the nCTEQ \cite{Kovarik:2015cma,Duwentaster:2022kpv,Muzakka:2022wey} and EPPS \cite{Eskola:2021nhw} parameterizations with those obtained neglecting the nuclear effects. Predictions for $\nu_{\tau} A$ and $\bar{\nu}_{\tau} A$ interactions will be presented considering different energies for the incoming tau neutrino and distinct scattering angles.

This paper is organized as follows. The next section presents a brief review of the formalism used to estimate the differential cross-sections and the longitudinal and transverse components of the tau lepton polarization. 
In Section \ref{Sec:Results} we present our results for these quantities considering  different neutrino energies and scattering angles. Finally, in Section \ref{Sec:conc}, we summarize our main results and conclusions.

\begin{figure}[t]
\includegraphics[scale=0.45]{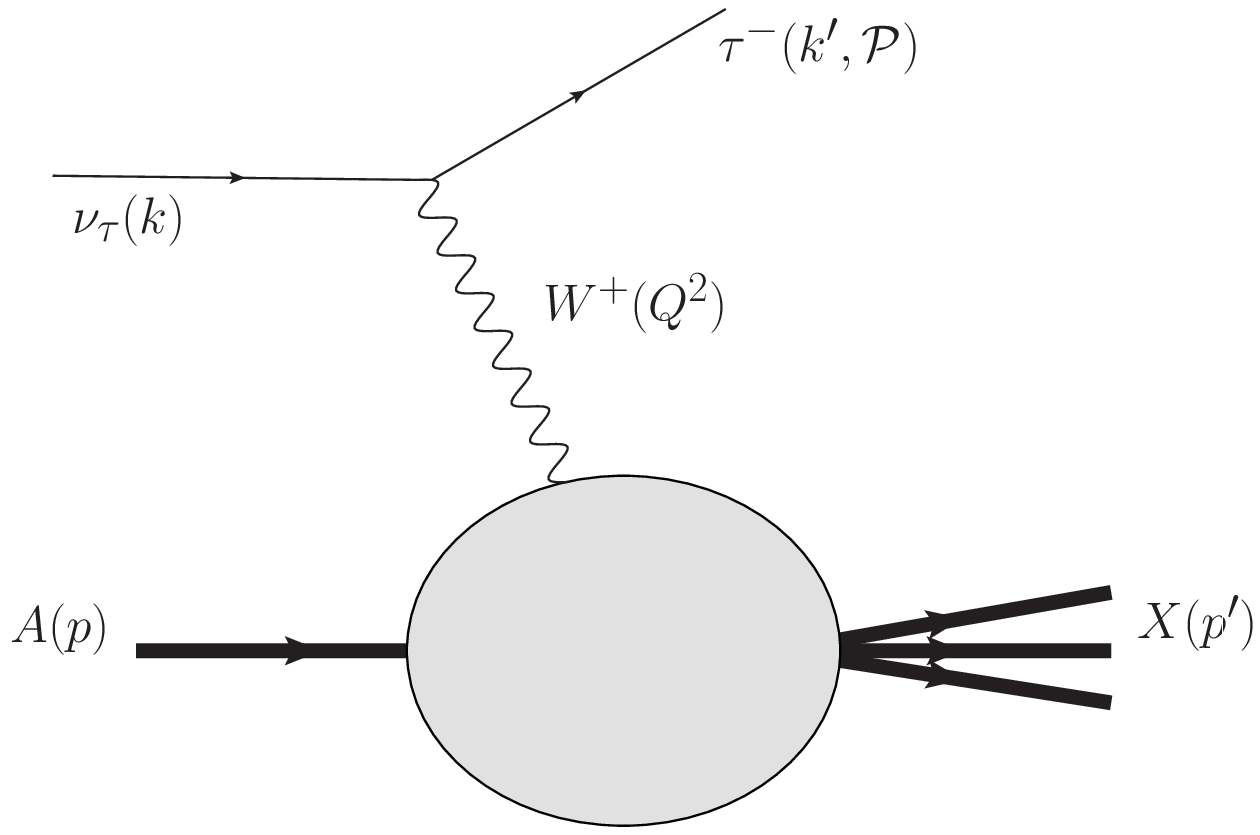} 
\caption{Production of a tau lepton with momentum $k^{\prime}$ and polarization ${\cal{P}}$ in a charged current (CC) deep inelastic  $\nu_{\tau} A$ scattering.}
\label{Fig:diagram}
\end{figure}

\section{Formalism}
In our analysis, we will investigate the degree of polarization $\cal{P}$ of a tau lepton produced in a charged current (CC) deep inelastic $\nu_{\tau} A$ scattering, represented in Fig. \ref{Fig:diagram},  in the laboratory frame. In this frame, one has that a tau neutrino with four-momentum $k^\mu = (E_\nu , 0, 0, E_\nu)$ collides with the nucleus target of four-momentum $p^\mu = (M_A , 0, 0, 0)$, becoming a tau of four-momentum $k'^{\mu} = (E_\tau, |\vec{k}'| \mathrm{sin}\, \theta, 0,  |\vec{k}'| \mathrm{cos}\, \theta)$. The interaction is mediated by the $W^{+}$ boson, characterized by a square four-momentum $Q^2 \equiv - q^2 = (k - k^{\prime})^2$ and mass $M_W$. The nucleus target  becomes an unknown final hadronic state $X$ with invariant mass $p^{\prime \, 2} = W^2 = (p + q)^2$. As usual, the cross-section for this process can be factorized in terms of the leptonic and hadronic tensors, which describe the upper and lower parts of the diagram presented in Fig. \ref{Fig:diagram}. For a polarized $\tau^{\pm}$ lepton, the leptonic tensor is given by \cite{Zaidi:2023hdd}
\begin{eqnarray}
    L^{pol}_{\mu\nu}(s,h) = \frac{1}{2} 
    [L^{unpol}_{\mu\nu}
    \mp h m_\tau s^{\alpha}
    (
    k_\mu g_{\nu\alpha} + k_{\nu}g_{\mu\alpha}
    -k_{\alpha} g_{\mu\nu} 
    \pm i \epsilon_{\mu\nu\alpha\beta} k^{\beta}
    )
    ] 
    \label{eq:tensorLpol}
\end{eqnarray}
where  
\begin{eqnarray}
    L^{unpol}_{\mu\nu} = 8(k_\mu k'_\nu + k_\nu k'_\mu - k\cdot k' g_{\mu \nu} \pm i\epsilon_{\mu\nu\rho\sigma} k^\rho k'^\sigma )
    \label{eq:tensorL}
\end{eqnarray}
and  $s^{\alpha}$ is the tau spin four-vector and $h=\pm 1$ the helicity (+ for neutrino and - for antineutrino). On the other hand, the hadronic tensor is expressed in terms of the nuclear structure functions, $F^A_{i}=F_{i}(x,Q^2)$, as follows
\begin{eqnarray}
\begin{aligned}
    W^{\mu\nu}_{A} = -g^{\mu \nu }F^A_{1}(x, Q^2) + 
    \frac{2x}{Q^2} p^{\mu} p^{\nu} F^A_{2}(x, Q^2) -
    \frac{i x}{Q^2}\epsilon^{\mu \nu\rho\sigma} p_{\rho} q_{\sigma} F^A_{3}(x, Q^2) + \\
    + \frac{2}{Q^2} q^{\mu} q^{\nu} F^A_{4}(x,Q^2) + 
    \frac{2x}{Q^2} (p^{\mu}q^{\nu} + q^{\mu}p^{\nu})F^A_{5}(x,Q^2) \, ,
    \label{eq:tensorW}
\end{aligned}
\end{eqnarray}
where  the Bjorken-$x$ variable is defined by $x = Q^2/(2p\cdot q)$. 

The polarized double differential cross-section, expressed in terms of the tau energy $E_{\tau}$ and the scattering angle $\theta$, is proportional to $L^{pol}_{\mu\nu}W^{\mu\nu}_{A}$ and is given by \cite{Zaidi:2023hdd}
\begin{eqnarray}
    \frac{\mathrm{d}^2\sigma_A^{pol}}{\mathrm{d}E_\tau\mathrm{d\,cos}\,\theta} = 
    \frac{1}{2}(1+s_\mu P^{\mu}) 
    \frac{\mathrm{d}^2\sigma_A}{\mathrm{d}E_\tau\mathrm{d\,cos}\,\theta}\, ,
    \label{eq:sigmaPol}
\end{eqnarray}
where $P^{\mu}$ is the tau polarization four-vector and  
\begin{eqnarray}
    \begin{aligned}
    \frac{\mathrm{d}^2\sigma_A}{\mathrm{d}E_\tau\mathrm{d\,cos}\,\theta} = 
    \frac{G_F^2|\vec{k}'|}{2\pi E_\nu (1+Q^2/M_W^2)^2} \left\{ 2F^A_{1}(x,Q^2)(E_\tau -|\vec{k}'|\mathrm{cos}\,\theta) + 
    F^A_{2}(x,Q^2)\frac{M_A}{\nu}(E_\tau+|\vec{k}'|\mathrm{cos}\,\theta) \right. \\
    \pm F^A_{3}(x,Q^2)\frac{1}{\nu}[| \vec{k}' |^2 + E_\nu E_\tau - (E_\nu + E_\tau )|\vec{k}'|\mathrm{cos}\,\theta] + 
    F^A_{4}(x,Q^2)\frac{m_\tau^{2}}{\nu M_A x} (E_\tau - |\vec{k}'|\mathrm{cos}\,\theta) + \\
    \left. - F^A_{5}(x,Q^2)\frac{2m_\tau^2}{\nu} \right\} \, ,
    \label{eq:sigma2}
    \end{aligned}
\end{eqnarray}
with the upper (lower) sign for (anti) neutrinos, $m_{\tau}$ the mass of the tau lepton and $\nu = E_{\nu} - E_{\tau}$.

The polarization vector $P^{\mu}$ can be decomposed in  terms of a longitudinal $P_L$ (in the direction of $k^{\prime}$), transverse $P_T$ (transverse to $k^{\prime}$ and contained in the neutrino - tau lepton plane) and perpendicular $P_P$(orthogonal direction to the neutrino - tau lepton plane) components. As demonstrated in Refs. \cite{Hagiwara:2003di,Hernandez:2022nmp,Zaidi:2023hdd,Isaacson:2023gwp}, the perpendicular component does not contribute, since the three-vector polarization lies in the direction perpendicular to the $\tau$ lepton scattering plane. In the laboratory frame, the relevant components of the polarization vector are given by \cite{Zaidi:2023hdd}
\begin{eqnarray}
    \begin{aligned}
    P_L = 
    \mp \frac{E_\nu}{L^{unpol}_{\mu\nu} W_{A}^{\mu\nu}} \left\{ \left[ 2F^A_{1}(x,Q^2) - F^A_{4}(x,Q^2)\frac{m_\tau^{2}}{\nu M_A x} \right] (|\vec{k}'| - E_\tau\mathrm{cos}\,\theta) + 
    F^A_{2}(x,Q^2)\frac{M_A}{\nu}(|\vec{k}'| - E_\tau\mathrm{cos}\,\theta) \right. \\
    \pm \frac{F^A_{3}(x,Q^2)}{\nu}\frac{1}{\nu}[ (E_\nu + E_\tau )|\vec{k}'| - (| \vec{k}' |^2 + E_\nu E_\tau)\mathrm{cos}\,\theta] + 
    \left. - F^A_{5}(x,Q^2)\frac{2m_\tau^2}{\nu} \mathrm{cos}\,\theta \right\}
    \label{eq:PL}
    \end{aligned}
\end{eqnarray}
and
\begin{eqnarray}
\begin{aligned}
    P_T = 
    \mp \frac{m_\tau \mathrm{sin}\,\theta E_\nu}{L^{unpol}_{\mu\nu} W_{A}^{\mu\nu}} \left[ 2F^A_{1}(x,Q^2) -  F^A_{2}(x,Q^2)\frac{M_A}{\nu} 
    \pm F^A_{3}(x,Q^2) \frac{E_\nu}{\nu} - F^A_{4}(x,Q^2)\frac{m_\tau^{2}}{\nu M_A x} 
     + F^A_{5}(x,Q^2)\frac{2E_\tau}{\nu} \right] \, .
     \label{eq:PT}
\end{aligned}
\end{eqnarray}
One has that $P_T$ is proportional to $\mathrm{sin}\,\theta$ and $m_\tau$, then vanishes for $\theta = 0^\circ$ and is enhanced for heavier leptons. Finally, the degree of polarization of the lepton produced is defined as
\begin{eqnarray}
    {\cal{P}} = \sqrt{P_L^2 + P_T^2} \, ,
    \label{eq:P}
\end{eqnarray}
 which is less than or equal to 1, being 1 when the particle is completely polarized and 0 unpolarized. 
 
 
  The main input in the calculations of the tau polarization are the nuclear structure functions, which can be expressed in terms of the nuclear parton distributions (nPDF) and are sensitive to the nuclear effects, with the magnitude of these effects being dependent on $x$, $Q^2$, parton specie and atomic mass number $A$ (For a recent review see Ref. \cite{Klasen:2023uqj}).   Over the last years, several groups have proposed parameterizations for these distributions, which are based on different assumptions and techniques to perform a global fit of different sets of data using the DGLAP evolution equations \cite{dglap}.  In our study, we will consider the nCTEQ15 \cite{Kovarik:2015cma,Duwentaster:2022kpv,Muzakka:2022wey} and EPPS21 \cite{Eskola:2021nhw} parametrizations, which make use of the Hessian method for the treatment of the propagation of experimental uncertainties into the nPDFs but differ in the assumptions for the initial condition of the DGLAP evolution equation as well as in the data sets used in the global fit. While in the nCTEQ15 framework, the nuclear PDFs are parametrized as the nucleon one, with the $A$ - dependence included in the coefficients of the parametrization, in the EPPS one, the nuclear effects are parametrized in a nuclear ratio in order to reduce the dependence on free proton PDFs. As shown in Ref.  \cite{Klasen:2023uqj}, both parametrizations provide a very good description of the current data for the observables measured in lepton - nucleus and proton - nucleus collisions. For completeness, we will compare the nCTEQ and EPPS predictions with those derived neglecting the nuclear effects. Additionally, in our analysis, we will assume the validity of the Callan-Gross and Albright-Jarlskog relations, which allow us to write $F^A_{1}$ and $F^A_{5}$ in terms of $F^A_{2}$, respectively.

\begin{figure}[t]
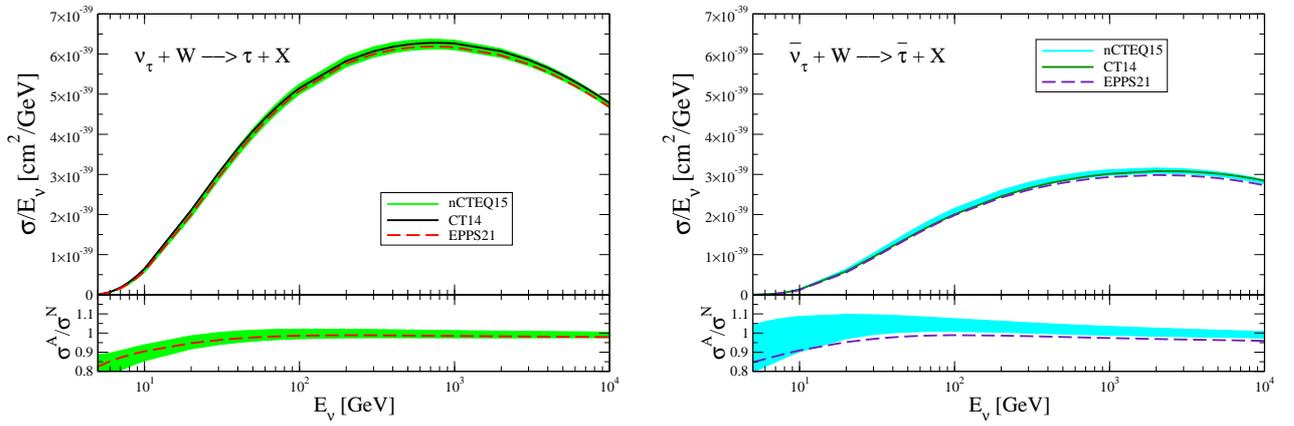

	\centering
	\begin{tabular}{ccc}
	\includegraphics[width=0.45\textwidth]{nu_tau_W_unpol.eps} & \,\,\,\,\, & \includegraphics[width=0.45\textwidth]{an_tau_W_unpol.eps} 
			\end{tabular}
\caption{Predictions for the dependence of the ${\nu}_{\tau}W$ (left panel)  and  $\bar{\nu}_{\tau}W$ (right panel) cross-sections on the energy of the incoming tau neutrino. Results derived assuming different parametrizations for the nPDFs. The predictions for the ratio between the nuclear and nucleon cross-sections are presented in the bottom panels.}
\label{fig:sigma}
\end{figure}

\section{Results}
\label{Sec:Results}

In this section, we will present our results for the total and differential cross-sections as well as for the longitudinal and transverse polarizations components considering the interaction of tau neutrinos and antineutrinos with a nuclear target. Motivated by the recent results obtained by the FASER$\nu$ experiment \cite{FASER:2023zcr}, we will assume a Tungsten target ($W$, with $A = 184$).   Initially, in Fig. \ref{fig:sigma} (left panel), we present our predictions for the  dependence of the total neutrino-tungsten cross-section per nucleon on the energy of the incoming tau neutrino, derived considering different parametrizations for the nPDFs. The corresponding predictions for the $\bar{\nu}_{\tau}W$ cross-section are presented in the right panel, assuming a distinct set of colors  for the different predictions in comparison with those used in the left panel, in order to help to distinguish the results for ${\nu}_{\tau}W$ from those for $\bar{\nu}_{\tau}W$ in what follows. 
Our predictions are derived assuming the nCTEQ15 and EPPS21 parametrizations for the nuclear PDFs. For the nCTEQ15 case, we present the associated uncertainty band. For the EPPS21 case, which parametrizes the ratios between nuclear and proton PDFs, $R_i(x,Q^2) = f_i^A (x,Q^2) / [A  f_i^p(x,Q^2)]$, we assume that the proton PDF is described by the CT14 parametrization \cite{ct14} and present only the central prediction.  In addition, we also present the results derived neglecting the nuclear effects, which were calculated assuming $R_i(x,Q^2) = 1$ ($\forall i$) and are denoted CT14 hereafter.
In the bottom panels of Fig. \ref{fig:sigma}  we show the ratio between the cross-sections with nuclear effects and those for a free nucleon. One has that $\sigma_{{\nu}_{\tau}W} > \sigma_{\bar{\nu}_{\tau}W}$ in the energy range considered and that the associated uncertainties vary between $5\%$ and $10\%$ of the total cross-section for the range $10^2 - 10^4$ GeV of incident tau neutrino energy. Moreover, the EPPS21 prediction for the ${\nu}_{\tau}W$ cross-section   is contained in the uncertainty band of the results obtained with nCTEQ15 parametrization. In contrast, for the $\bar{\nu}_{\tau}W$ cross-section, the EPPS21 central prediction implies a smaller cross-section than the nCTEQ15 band, which is associated with the larger amount of shadowing predicted by this parametrization in comparison with the nCTEQ15 one. However, one has verified that if the uncertainties on the EPPS21 parametrization are taken into account, the associated band overlaps with the nCTEQ15 one.

\begin{figure}[t]
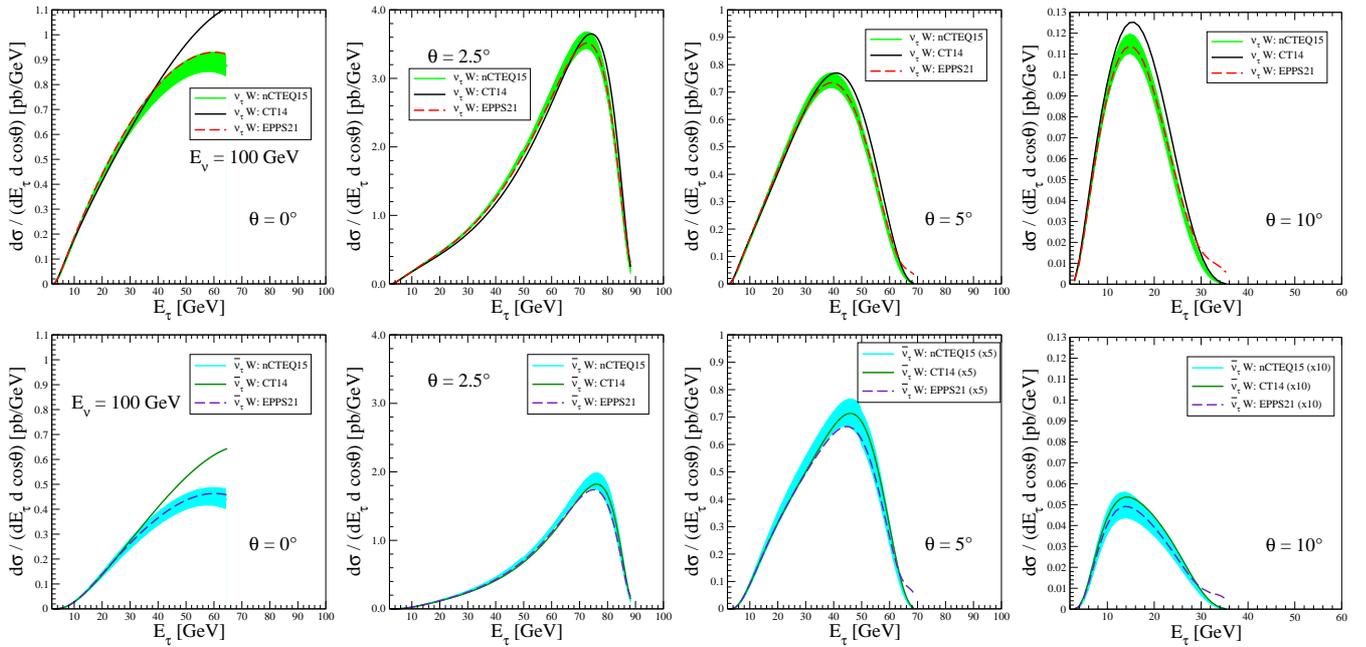

	\centering
	\begin{tabular}{ccccccc}
     \includegraphics[width=0.24\textwidth]{dsigdEdc_nu_0graus_E_100.eps} &            \includegraphics[width=0.24\textwidth]{dsigdEdc_nu_2.5graus_E_100.eps} & 
     \includegraphics[width=0.24\textwidth]{dsigdEdc_nu_5graus_E_100.eps} & \includegraphics[width=0.24\textwidth]{dsigdEdc_nu_10graus_E_100.eps}          \\
    \includegraphics[width=0.24\textwidth]{dsigdEdc_an_0graus_E_100.eps} &  \includegraphics[width=0.24\textwidth]{dsigdEdc_an_2.5graus_E_100.eps} &  
     \includegraphics[width=0.24\textwidth]{dsigdEdc_an_5graus_E_100.eps} &  \includegraphics[width=0.24\textwidth]{dsigdEdc_an_10graus_E_100.eps}         
			\end{tabular}
\caption{ Double differential ${\nu}_{\tau}W$ (upper panels)  and  $\bar{\nu}_{\tau}W$ (lower panels) cross-sections as a function of the tau lepton energy for different values of the 
angle $\theta$. Results derived assuming different nPDFS and that the energy of the incoming tau neutrino is equal to 100 GeV. Note the different $y$-axis scales in the distinct plots. }
\label{fig:dsdE_nu100}
\end{figure}

\begin{figure}[t]
	\centering
	\begin{tabular}{ccccccc}
    \includegraphics[width=0.24\textwidth]{dsigdEdc_nu_0graus_E_1000.eps} & \includegraphics[width=0.24\textwidth]{dsigdEdc_nu_2.5graus_E_1000.eps} &
    \includegraphics[width=0.24\textwidth]{dsigdEdc_nu_5graus_E_1000.eps} & \includegraphics[width=0.24\textwidth]{dsigdEdc_nu_10graus_E_1000.eps}  \\
    \includegraphics[width=0.24\textwidth]{dsigdEdc_an_0graus_E_1000.eps} &  \includegraphics[width=0.24\textwidth]{dsigdEdc_an_2.5graus_E_1000.eps} & 
    \includegraphics[width=0.24\textwidth]{dsigdEdc_an_5graus_E_1000.eps} &  \includegraphics[width=0.24\textwidth]{dsigdEdc_an_10graus_E_1000.eps}    
			\end{tabular}
\caption{ Double differential ${\nu}_{\tau}W$ (upper panels)  and  $\bar{\nu}_{\tau}W$ (lower panels) cross-sections as a function of the tau lepton energy for different values of the 
angle $\theta$. Results derived assuming different nPDFS and that the energy of the incoming tau neutrino is equal to 1000 GeV. Note the different $y$-axis scales in the distinct plots.}
\label{fig:dsdE_nu1000}
\end{figure}


In Figs. \ref{fig:dsdE_nu100} and \ref{fig:dsdE_nu1000} our predictions for the double differential ${\nu}_{\tau}W$ (upper panels)  and  $\bar{\nu}_{\tau}W$ (lower panels) cross-sections as a function of the tau lepton energy for different values of the 
angle $\theta$, which is defined as the scattering angle of tau in the laboratory frame in relation to the incoming tau (anti) neutrino axis. Results derived assuming different nPDFS and that the energy of the incoming tau (anti) neutrino is equal to 100 GeV in Fig. \ref{fig:dsdE_nu100} and 1000 GeV in Fig. \ref{fig:dsdE_nu1000}.
One has that the differential distributions for neutrinos are larger than for antineutrinos in all kinematic conditions presented. For $E_{\nu} = 100$ GeV (Fig. \ref{fig:dsdE_nu100}),  the cross-section is larger for $\theta = 2.5^\circ$ than for $\theta = 0^\circ$, but decreases for larger angles ($5^\circ$ and $10^\circ$). This behavior does not occur for $E_{\nu} = 1000$ GeV (Fig. \ref{fig:dsdE_nu1000}), where we as  that the differential distribution is greater for $\theta = 0^\circ$ than for the other angles presented. Our results indicate that,  increasing the energy of the incident neutrino, the distribution is shifted towards small values of $\theta$. Furthermore, one has that for higher scattering angles of the produced tau, the typical tau energy becomes increasingly smaller, and values close to the neutrino energy are not allowed. For example, for neutrinos with a energy of 1000 GeV, the tau energy is lower than 660 GeV, 500 GeV, 200 GeV and 56 GeV for $\theta = 0^\circ$, $2.5^\circ$, $5^\circ$ and $10^\circ$, respectively. 
Regarding the treatment of the nuclear effects, one has that the nCTEQ15 and EPPS21 predictions are similar. In contrast, the CT14 prediction, which disregard the  nuclear effects, implies results similar to those obtained using the nCTEQ15 and EPPS21 parametrizations for the  angles of $2.5^\circ$, $5^\circ$ and $10^\circ$, but differs significantly for $\theta = 0^{\circ}$. Comparing the predictions of nCTEQ15 and CT14 for $\theta = 0^\circ$ and the maximum value allowed for $E_\tau$, one has an increasing of $153\%$ ($191\%$) when we neglect the nuclear effects for $E_{\nu} = 100$ (1000) GeV.

\begin{figure}[t]
	\centering
	\begin{tabular}{ccccc}       
 \raisebox{0.2\height}{\includegraphics[width=0.2\textwidth]{legenda.eps}} &
 \includegraphics[width=0.24\textwidth]{PT_2.5graus_E_100.eps} & 
 \includegraphics[width=0.24\textwidth]{PT_5graus_E_100.eps} &  \includegraphics[width=0.24\textwidth]{PT_10graus_E_100.eps} \\
 \includegraphics[width=0.24\textwidth]{PL_0graus_E_100.eps} & \includegraphics[width=0.24\textwidth]{PL_2.5graus_E_100.eps} & 
 \includegraphics[width=0.24\textwidth]{PL_5graus_E_100.eps} &  \includegraphics[width=0.24\textwidth]{PL_10graus_E_100.eps}
			\end{tabular}
\caption{ Transverse (upper panels) and longitudinal (lower panels) components of the polarization vector as a function of the tau lepton energy for an incident  (anti) neutrino with an energy of 100 GeV. Results for  different values for the angle $\theta$ derived assuming distinct nPDFs.}
\label{fig:PL100}
\end{figure}

\begin{figure}[t]
	\centering
	\begin{tabular}{ccccc} 
    \raisebox{0.2\height}{\includegraphics[width=0.2\textwidth]{legenda.eps}} &
    \includegraphics[width=0.24\textwidth]{PT_2.5graus_E_1000.eps} &  
    \includegraphics[width=0.24\textwidth]{PT_5graus_E_1000.eps} &  \includegraphics[width=0.24\textwidth]{PT_10graus_E_1000.eps} \\
    \includegraphics[width=0.24\textwidth]{PL_0graus_E_1000.eps} &
    \includegraphics[width=0.24\textwidth]{PL_2.5graus_E_1000.eps} &  
    \includegraphics[width=0.24\textwidth]{PL_5graus_E_1000.eps} &  \includegraphics[width=0.24\textwidth]{PL_10graus_E_1000.eps}
			\end{tabular}
\caption{ Transverse (upper panels) and longitudinal (lower panels) components of the polarization vector as a function of the tau lepton energy for an incident  (anti) neutrino with an energy of 1000 GeV. Results for  different values for the angle $\theta$ derived assuming distinct nPDFs. }
\label{fig:PL1000}
\end{figure}

In Figs.  \ref{fig:PL100} and \ref{fig:PL1000} we present our results for the  transverse ($P_T$, upper panels) and longitudinal ($P_L$, lower panels) components of polarization as a function of the tau lepton energy and different values of $\theta$, for an incident neutrino energy of 100 GeV and 1000 GeV, respectively. 
 For $\theta = 0^\circ$ we have $P_L$ almost constant and close to 1 for neutrinos and -1 for antineutrinos. For the other angles, we have that $P_L$ is positive (negative) for low values of tau (antitau) energy and becomes negative (positive) at higher energies (close to the maximum allowed energy). The behavior of the curves is similar when increasing the energy of the incident neutrino, differing basically in the energy range allowed for tau. Using different parametrizations for the nPDFs we obtain similar predictions  for $P_L$ in all kinematic conditions presented in Figs. \ref{fig:PL100} and \ref{fig:PL1000}, where one has  the overlapping of the distinct curves in practically all cases. 
Regarding the behavior of $P_T$, one has that for $\theta = 0^\circ$,  $P_T$ is equal to zero, which is expected since $P_T \propto \mathrm{sin}\,\theta$. For the other angles, $P_T$ grows with the tau energy until a maximum and then decreases to a value close to zero. In ${\nu}_{\tau}W$ interactions, $P_T$ presents only negative values, while for antineutrinos only positive values are observed. The predictions derived assuming distinct nPDFs are very similar.
In contrast, for $\bar{{\nu}}_{\tau}W$ interactions, the predictions differ for larger scattering angles. It is important to emphasize that the impact of the uncertainties present in the nCTEQ15 predictions becomes very small in these observables, which is directly associated with its definitions, Eqs. (\ref{eq:PL}) and (\ref{eq:PT}), which are given in terms of ratios of nuclear PDFs.

\begin{figure}[t]
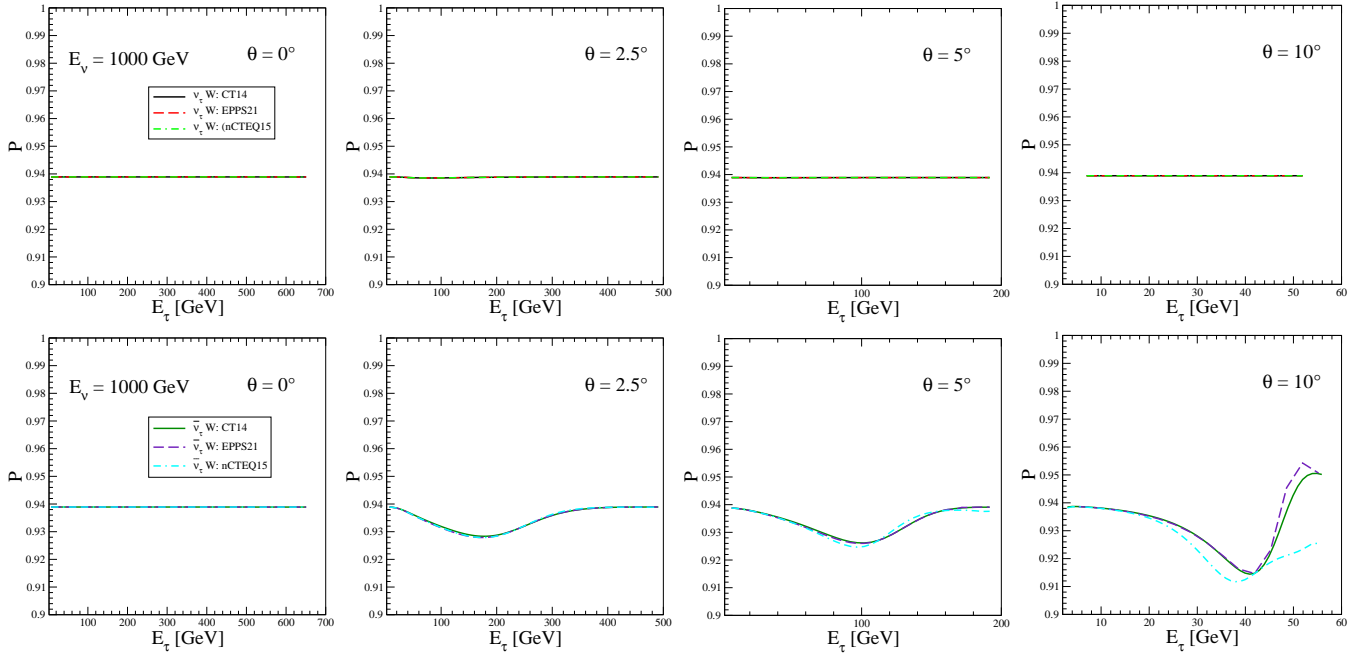

	\centering
	\begin{tabular}{ccccccc}
    \includegraphics[width=0.24\textwidth]{P_0graus_nu_E_1000.eps} & \includegraphics[width=0.24\textwidth]{P_2.5graus_nu_E_1000.eps} &  
    \includegraphics[width=0.24\textwidth]{P_5graus_nu_E_1000.eps} &  \includegraphics[width=0.24\textwidth]{P_10graus_nu_E_1000.eps}          \\
    \includegraphics[width=0.24\textwidth]{P_0graus_an_E_1000.eps} &  \includegraphics[width=0.24\textwidth]{P_2.5graus_an_E_1000.eps} &  
    \includegraphics[width=0.24\textwidth]{P_5graus_an_E_1000.eps} &  \includegraphics[width=0.24\textwidth]{P_10graus_an_E_1000.eps} 
			\end{tabular}
\caption{ Degree of polarization of the  tau (upper panels) and antitau (lower panels) produced in  ${\nu}_{\tau}W$   and  $\bar{\nu}_{\tau}W$ interactions, respectively,  as a function of the tau lepton energy for an incident  (anti) neutrino with an energy of 1000 GeV. Results for  different values for the angle $\theta$ derived assuming distinct nPDFs.  }
\label{fig:P}
\end{figure}

Finally, in Fig. \ref{fig:P} we present our results for  the degree of polarization $\cal{P}$ ($= \sqrt{P_L^2 + P_T^2}$) of taus (upper panels) and antitaus (lower panels) produced in  ${\nu}_{\tau}W$   and  $\bar{\nu}_{\tau}W$ interactions, respectively,  as a function of the tau lepton energy for an incident  (anti) neutrino with an energy of 1000 GeV. Similar results are derived for $E_{\nu} = 100$ GeV and are available upon request. One has that for ${\nu}_{\tau}W$  interactions, 
$\cal{P}$ $\approx 0.94$, with the predictions being almost independent of the scattering angle $\theta$, tau lepton energy and the nPDF used as input in the calculations. In contrast, for $\bar{\nu}_{\tau}W$ interactions, the degree of polarization is dependent on $\theta$, with the predictions for $\theta > 0^\circ$ presenting a minimum in the tau lepton energy range  considered. In addition, for this case, the predictions for large angles become dependent on the nPDF considered.


\section{Summary}
\label{Sec:conc}
In recent years, the study of neutrino - nucleus interactions with incident neutrino energy ranges in the GeV -- TeV range becomes a reality with the first measurements performed in the LHC by the FASER  and SND@LHC experiments and a larger amount of data is expected in the forthcoming years with the construction of the Forward Physics Facility. Such data will allow us to study in more detail the  tau neutrino properties, its interaction and the decay of the produced tau lepton. In this paper, we have investigated the degree of polarization of the (anti) tau lepton produced in (anti) neutrino - tungsten interactions, considering different values for the energy of the incident (anti) neutrino and distinct parametrizations for the treatment of the nuclear effects in the parton distribution functions.  
We have reviewed the formalism and presented predictions for the  differential cross-sections and the longitudinal and transverse components of the tau lepton polarization as a function of the tau lepton energy and distinct values of the scattering angle. Our results indicate that the degree of polarization is smaller than 1 for the neutrino energies reached at the LHC and are almost insensitive to the nuclear effects.

\begin{acknowledgments}

R. F. acknowledges support from the Conselho Nacional de Desenvolvimento Científico e Tecnológico (CNPq, Brazil), Grant No. 161770/2022-3. V.P.G. was partially supported by CNPq, FAPERGS and INCT-FNA (Process No. 464898/2014-5).
 D.R.G. was partially supported by CNPq.
 
\end{acknowledgments}

\hspace{1.0cm}

\end{document}